\input psfig
\documentstyle{article}
\def\vol{\frac{d^2 k}{(2\pi)^2}}
\def\val{\frac{d^3 k}{(2\pi)^3}}

\def\beq{\begin{equation}}
\def\eeq{\end{equation}}
\def\bqq{\begin{eqnarray}}
\def\eqq{\end{eqnarray}}

\def\ks{k \!\!\! /}
\def\qs{q \!\!\! /}
\def\ps{p \!\!\! /}

\def\vertical{\left. \right|}
\def\dag{^{\dagger}}
\begin{document}
\vspace{24pt}
\begin{center}
{\bf \Large  Finite Temperature and Density Effects in Planar Q.E.D}
\baselineskip=12pt
\vspace{35pt}

Gil Gat$^{1}$ and Rashmi Ray$^2$
\vspace{24pt}

Physics Department\\
University of Maryland\\
College Park, MD 20742\\
U.S.A\\
\vspace{130pt}
\end{center}
\begin{abstract}
The behavior of finite temperature planar electrodynamics is investigated. We
calculate the static as well as dynamic characteristic functions  using
real time formalism. The temperature and density dependence of
dielectric and permeability functions, plasmon frequencies and their
relation to the screening length is determined. The radiative correction
 to the
fermion mass is also calculated. We also calculate the
 temperature dependence of the
electron (anyon) magnetic moment. Our results for the gyromagnetic ratio
go smoothly to the known result at zero temperature, $g=2$, in accordance with
the general expectation.
\end{abstract}
\vspace{50pt}
$^1$ggat@delphi.umd.edu
\newline
$^2$rray@delphi.umd.edu

\vfill
\baselineskip=20pt
\pagebreak

{\bf Introduction}

Perturbative quantum field theory at finite temperature and density is
of current interest in many areas of elementary particle physics. Apart
from its cosmological relevance it is important in determining the
properties of systems like the quark-gluon plasma in Q.C.D and the
electron-positron plasma in Q.E.D. It is also of some interest in a
 more exact
description of nuclear matter \cite{bech}. The recent interest in planar
 systems
motivates the investigation of such sytems in 2+1 dimensions \cite{baruch};
  further
since by dimensional arguments the behavior of many physical quantities
at low temperature is expected to be enhanced in 2+1 dimensions
 (i.e. $\sim T$ instead of
$\sim T^2$ in 3+1 dimensions) it might provide a simpler way of
testing Q.E.D in 2+1 accurately.

 In this paper we investigate some
properties of the two and three point functions in 2+1 dimensional Q.E.D.
The Green functions are the basic tools in the study of quantum many
body systems. Knowledge of the Green functions allows one to calculate
all  physically interesting quantities like energy, density, thermodynamic
potential etc. Furthermore, the singularities in momentum space
of these functions determine the spectrum of excitations of the system.
 Linear response theory  relates the Green functions to the various
quantities that characterize the response of the system to small external
fields. In the following sections we use the real time formalism
\cite{kap} to
calculate the polarization tensor, the fermion self energy and the vertex
function at finite density and temperature. We calculate static
properties like the dielectric constant and magnetic permeability as well as
the magnetic and electric screening lengths. The magnetic mass
vanishes in the leading order as in the 3+1 dimensional case.
Dynamical properties such as the  transverse and
longitudinal plasmon frequency can also be extracted
from the photon two point function. We find that for small momenta
 $(m_{pl}^{L})^2 =( m_{pl}^{T})^2=\frac{1}{2}m_{el}^2$
(in the 3+1 dimensional case , for small momenta $(m_{pl}^{L})^2=
(m_{pl}^{T})^2=\frac{1}{3}m_{el}^2$).

In 2+1 dimensions there is the possibility of adding a topological gauge
 invariant mass term , the Chern Simons term (CS), to the action of Q.E.D .
The spin and statistics of electrons in this case is altered by this term,
$S=\frac{1}{2}+\frac{1}{\kappa}$ where $\kappa$ is the CS coefficient.
 The finite temperature behavior of the CS term has been investigated by
several authors \cite{aitch},\cite{das},\cite{pani}. Here we calculate
its dependence on the density and
temperature.  The spin of electrons (anyons) is related to the
anomalous magnetic moment by the $g$ factor, $\mu=\frac{g}{2m}S$. It has
 been observed \cite{nair} that at zero temperature $g=2$ and that this
 result is an exact one. This was verified to one loop in perturbation
 theory \cite{kogan}. From the three point function we can directly calculate
the anomalous magnetic moment at finite temperature. We
verify that to one loop and in the limit $T, \mu \rightarrow 0$, $g=2$.

{\bf Finite Temperature Propagators}

There exist two main formulations of field theory at finite temperature.
In the imaginary time formalism one replaces the continuous energy
variable $k_0$ by $2\pi i n T $ for bosons and $2\pi i(n+{1\over 2})T$
for fermions. and the integration over this variable by a discrete
sum $\int {{dk}\over{2\pi }}\rightarrow iT
\sum_{\rm n=-\infty }^{\infty} $.
{}From this one immediately notices that in the very high temperature case
the time direction disappears entirely because the temporal interval
shrinks to zero. Also, in this limit only the zero energy processes
survive, as nonzero energy necessarily means high energy at high
temperatures, owing to discrete nature of the energy variable. Thus
at high temperatures all Matsubara modes with $n\neq 0$ decouple.
This feature is certainly attractive in a study of static processes.
However, one might want to study processes where the energy is neither
of two extremes--large or zero. The imaginary time formalism is clearly
not ideally suited for this regime.
Apart from this, if one is interested in addressing non-static problems
one has to analytically continue the results in order to recover the real
time. This process of analytic continuation can actually become quite
arduous in many cases of interest.
In the real time formulation, on the other hand the energies
remain real but one has to accept a doubling of the number of degrees
of freedom. Time in this formalism is treated as a complex quantity
and the real time is actually just one branch (A) of a time contour
(Fig. 1) in the
complex plane, that lies on the real axis. The other branches of the
contour however necessitate the introduction of unphysical degrees
of freedom which have support on the branch (B) of the contour.
Consequently, we are required to deal with a minimum of four different
propagators between AA, BB, AB and BA respectively \cite{weert}.

Here we just quote the expressions for the fermion and the photon
propagators of the AA kind.
The fermion propagator is
\beq
S_F(k)=\frac{i}{\ks -m +i\epsilon}-2 \pi (\ks+m)\delta (k^2-m^2)n_F(k_0)
\, \, ,
\eeq
where
\beq
n_F(k_0)=\frac{\Theta (k_0)}{e^{\beta (k_0 + \mu)}+1} +
\frac{\Theta (-k_0)}{e^{\beta (k_0 - \mu)}+1}  \, \, .
\eeq
The photon propagator in 2+1 dimensions in the Landau gauge is
\beq
D_{\mu \nu}(k)=-i \left[ g_{\mu \nu}-\frac{k_{\mu}k_{\nu}}{k^2}+iM
 \epsilon_{\mu
\nu \rho} \frac{k^{\rho}}{k^2} \right] \left(\frac{1}{k^2-M^2}-2 \pi i
\delta(k^2-M^2) n_B(k_0) \right) \, ,
\eeq
where
\beq
n_B(k_0)=\frac{1}{e^{\beta k_0}-1}  \, \, .
\eeq
In one loop calculations we need only these propagators \cite{pani}.

{\bf Photonic Two point function}

The polarization tensor at finite temperature and density determines
the dielectric constant , magnetic permeability  and other physically
interesting quantities \cite{weld}. Here we calculate the vacuum polarization
to one loop at finite temperature and density $\mu$.
The usual notion of temperature presupposes the existence of a preferred
frame of reference. We choose to stay with this definition and thus lose a
manifestly covariant description. We  therefore compute $\Pi_{00}$
and $\Pi_{ij}$ separately \footnote{of course one could formulate the problem
covariantly
{ \it a la} Weldon \cite{weld}.}.
  In contrast to similar
calculations in 3+1 dimensions it is possible to express all the results
in terms of simple analytic functions.

The most general expression for the parity-even part of the photon
two point function is given by \cite{weld}
\beq
\Pi_{\mu \nu }\equiv \Pi_T(\omega ,\vec k)P_{\mu \nu }+ \Pi_L(\omega ,\vec
k)Q_{\mu \nu } \, \, ,
\eeq
where
\beq
P_{\mu \nu }\equiv g_{\mu \nu }+{{k_{\mu }k_{\nu }}\over{{\vec k}^2}}
-{{k_0}\over{{\vec k}^2}}(k_{\mu }g_{\nu 0}+k_{\nu }g_{\mu 0})
+{{k^2}\over{{\vec k}^2}}g_{\mu 0}g_{\nu 0}
\eeq
and
\beq
Q_{\mu \nu }\equiv -{1\over{k^2{{\vec k}^2}}}(k_0 k_{\mu }-k^2 g_{\mu 0})
(k_0 k_{\nu }-k^2 g_{\nu 0})   \, \, .
\eeq
This satisfies the Ward identity
\beq
k^{\mu }\Pi_{\mu \nu }=0  \, \, .
\eeq
The functions $\Pi_T $ and $\Pi_L $ can be obtained from $\Pi_{\mu \nu }$
as
\bqq
\Pi_L(\omega ,{\bf k})&=& -{{k^2}\over{{\vertical {\bf k} \vertical
}^2}}\Pi_{00}
(\omega ,\bf k) \, \, , \nonumber \\
\Pi_T(\omega ,{\bf k})&=& {1\over 2}g^{ij}\Pi_{ij}(\omega ,{\bf k}) \, \, .
\label{pi}
\eqq
We are mostly concerned with the static limit, where the time component
$k_0$ of the four-vector $k_{\mu }$ is set to zero. This is appropriate
to the case where the linear response to a static external
electromagnetic field is investigated. Here, the Ward identity assumes
the form
\beq
k^i \Pi_{i \nu }=0 \, \, .
\eeq
In this particular situation, where $P_{i0}=Q_{i0}=Q_{ij}=0$, this is
tantamount to
\beq
k^i \Pi_{ij}=0   \, \, .
\eeq
Thus
\beq
\Pi_{ij}= \Pi_T(0,\vec k)(g_{ij}+{{k_i k_j}\over{{\vec k}^2}})  \, \, .
\eeq
It is easily verified that the trace of the polarization tensor is
just $\Pi_T(0,\vec k)$ in this case.

In the real time formalism one can easily separate the vacuum
contribution from the finite density and temperature contribution. In the
leading order $\Pi_{\mu \nu}(p)$ can be written as
\bqq
\Pi_{\mu \nu}(p)&=&i e^2 \int \frac{d^3k}{(2 \pi)^3} \, \, Tr[ \gamma_{\mu}
S_F (p+k)\gamma_{\nu} S_F (k)] \\ \nonumber
&=&\Pi_{\mu \nu}^{(0,0)} (p)+\Pi_{\mu \nu}^{(\beta, \mu)}(p)   \, \, ,
\eqq
where $\Pi_{\mu \nu}^{(0,0)}(p)$ is the zero temperature and density
polarization tensor. The temperature and density dependent terms
are given by:
\bqq
\Pi_{\mu \nu}^{(\mu,\beta)}(p) & =& ie^2 \int  \frac{d^3 k}{(2 \pi)^3}
\, L_{\mu \nu}(p,k) 2\pi i \left\{ \frac{n_{F}(k) \delta (k^2-m^2)}{
(k+p)^2-m^2+i\epsilon}+\frac{n_{F}(p+k)\delta ((p+k)^2-m^2)}{
k^2-m^2+i\epsilon } \right.
\nonumber \\
&+& \left. 2\pi i \, n_F(k) n_F(k+p) \delta(k^2-m^2) \delta ((k+p)^2-m^2)
 \right\}  \, \, ,
\eqq
where
\beq
L_{\mu \nu}(p,k)=2[2 k_{\mu} k_{\nu}+p_{\nu} k_{\mu}+p_{\mu} k_{\nu}-
g_{\mu \nu} (k^2-m^2+p \cdot k)+im \epsilon_{\mu \nu \rho} p^{\rho}] \, \, .
\eeq
{}From this rather general expression for the one-loop polarization tensor,
we extract useful information about the screening length of the electric
field at finite temperature and density ( from the electric mass $m_{el}$),
the static dielectric function and the static magnetic susceptibility.

If the electric field due to a charge placed in a medium picks up a mass
term as a result of interactions with the medium, it becomes short ranged
and the charge is screened. Screening occurs naturally
in a plasma as the charge attracts a halo of charges of the opposite
signature around it, polarizing the plasma and thereby effectively reducing
its own strength. The starting point for a calculation of the screening
length is the vacuum polarization (photonic two point) function
in the plasma in the static limit.
In fact, the potential energy of a static $e^{+}$ - $e^{-}$ pair ,
$ V({\bf R}, \mu ,T)$
is given by
\bqq
V({\bf R}, \mu , T) &=& \int \vol e^{i {\bf k}\cdot {\bf R}} V({\bf k},
\mu , T)  \nonumber \\
&=& -\int \vol e^{i{\bf k} \cdot R} \frac{e^2}{{\bf k}^2} [1-\Pi_{00}(0,{\bf
k})/{\bf k}^2]
 \, \, .
\eqq
For large ${\bf R}$ this potential becomes $V({\bf R}, \mu , T) \rightarrow
\frac{e^2}{4 \pi} {{e^{-m_{el}{\bf R}}}\over{\sqrt {m_{el}{\bf R}}}}$ , so
that it is actually screened with screening length $\simeq {1\over{m_{el}}}$.
In order to
calculate the screening length of static external electric fields
we compute $\Pi_{00}$ in the static infra red limit
$\lim_{{\bf p} \rightarrow 0} p^{\mu}=(0,{ \bf p})$:
\beq
Re (\Pi_{00}({\bf p})) = m_{el}^2 + \tilde{\Pi}_{00}({\bf 0}) {\bf p}^2+ \ldots
\eeq
We thus first  restrict our attention to the real part of $\Pi_{00}$.
{}From eq.(14) it follows that the terms that contribute to this part are
linear
in $n_F$.
\beq
Re (\Pi_{00}^{(\mu , \beta)}({\bf p}))= -e^2 \int \frac{d^3 k}{2 \pi^2}
(k_0^2+w^2+{\bf p} \cdot {\bf k})\left[ \frac{\delta(k_0^2-w^2)}{k_0^2
-{\bf p}^2
-2 {\bf p} \cdot {\bf k}-w^2}+   \frac{\delta(k_0^2-w'^2)}{k_0^2-w^2}
 \right] n_f(k)
\label{p00}
\eeq
where $w=\sqrt{ {\bf k}^2+m^2}$ and $w'=\sqrt{w^2+{\bf p}^2+2 {\bf p} \cdot
{\bf k}}$
. Expanding the second term in eq.(\ref{p00}) around ${\bf p}=0$ we find that
it
exactly cancels the
contribution of the first term, with the remaining contribution being
easy to evaluate through integration by parts.
Thus:
\beq
m_{el}^{2}=\frac{e^2 }{2 \pi \beta} \log [1+2e^{-\beta m} \cosh (\mu
 \beta)+e^{-2
\beta m}] +\frac{e^2 m}{2 \pi} \left[ \frac{1}{e^{\beta (m-\mu)}+1}+
\frac{1}{e^{\beta (m+\mu)}+1} \right]  \, \, ,
\label{elmass}
\eeq
and
\beq
\tilde{\Pi}_{00}({\bf 0}) =-\frac{e^2}{2 \pi m} \left[ \frac{1}{e^{\beta
(m+\mu)}
+1} +\frac{1}{e^{\beta (m-\mu)}+1} \right]-\frac{\beta e^2}{8 \pi} \left[
\frac{e^{\beta(m+\mu)}}{(e^{\beta (m+\mu)}+1)^2} +\frac{e^{\beta(m-\mu)}}
{(e^{\beta (m-\mu)}+1)^2} \right]  \, \, .
\eeq
In the limit $ T > \mu >m$, $m_{el}^2$ becomes $m_{el}^2 \simeq \frac{e^2
T}{\pi}
\log (\mbox{\small 2}) $. For $\mu > T > m$  we find $m_{el}^2 \simeq \frac{e^2
m}{2 \pi}
[1+e^{-\beta (m+\mu)}] $ (see fig. \ref{fig2}, \ref{fig3}).

The dielectric function which describes the response of the system to
small constant external electric fields is given by
\beq
Re\ {\bf \epsilon } ({\bf p}) =1+\frac{\Pi_{00}(0,{\bf p})}{{\bf
p}^2}=1+\tilde{\Pi}_{00}
(0,{\bf 0})+\frac{m_{el}^2}{{\bf p}^2}  \, \, .
\eeq

The spatial part of the polarization tensor has the following form in the
static limit
\beq
\Pi_{ij}=\Pi^{T} \left( g_{ij}+\frac{{\bf p}_i {\bf p}_j}{{\bf p}^2} \right)
\eeq
Taking the trace we get $\Pi^{i}_{\, i}(0,{\bf p})= {\bf p}^2
 \Pi^{T}({\bf p}^2)$ or
\beq
Re (\Pi^{i}_{\, i}({\bf p}))=-e^2 \int \frac{d^3 k}{\pi^2}(w^2-k_0^2-{\bf k}^2)
\left\{ \frac{\delta (k_0^2-w^2)}{k_0^2-w'^2}+\frac{\delta (k_0^2-
w'^2)}{k_0^2-w^2}
\right\}n_F(k)  \, \, .
\eeq
Expanding $w'$ and collecting all terms we find
\beq
Re(\Pi^{T}(0,{\bf p}))=\frac{3e^2}{4 \pi m} \left[\frac{1}{e^{\beta
(m+\mu)}+1}+
\frac{1}{e^{\beta (m-\mu)}+1} \right] {\bf p}^2+O({\bf p}^4)
\, \, .
\eeq
This can now be used to determine the magnetic permeability function
 $\mu(0, {\bf p})$  which is defined by (see Fig. \ref{fig4})
\beq
\frac{1}{\mu (0,{\bf p})}=1+\frac{Re \Pi^{T}(0,{\bf p})}{{\bf p}^2}=1+
\frac{3e^2}{4 \pi m} \left[\frac{1}{e^{\beta (m+\mu)}+1}+
\frac{1}{e^{\beta (m-\mu)}+1} \right] +O({\bf p}^2) \, \, .
\eeq
We note that as expected $\epsilon(0 ,{\bf p})\mu(0,{\bf p})=1$ for
$T \ll m$.

The parity odd part of the polarization tensor is especially interesting
in 2+1 dimensions. In particular the parity odd part of $\Pi_{\mu \nu}$ is
known to generate a topological gauge invariant mass \cite{jacktemp}.
Following
 \cite{pani} we write the full real time fermion propagator as
\beq
S_F^{\mu ,\beta}(p)=U_F(p_0)S_F^{0,0}(p)U_F(p_0)^{\dagger} \, \, ,
\eeq
where
\beq
U_F(p_0)= \left[ \begin{array}{cc}
       \cos \Theta (p, \beta \mu) & -e^{\beta \mu/2} \sin \Theta (p,
\beta \mu)  \\
     e^{-\beta \mu/2} \sin \Theta (p, \beta \mu )&  \cos \Theta (p,
\beta \mu)
     \end{array}
     \right]
\eeq
and
\bqq
\ cos \Theta (p, \beta \mu ) &=& \frac{ \Theta(p_0)e^{\beta(p_0-\mu )/4}
+\Theta(-p_0)e^{-\beta(p_0-\mu )/4}}{\sqrt{e^{\beta(p_0-\mu )/2}+
e^{-\beta(p_0-\mu )/2}}}  \\ \nonumber
\sin \Theta (p, \beta \mu ) &=& \frac{ \Theta(p_0)e^{\beta(p_0-\mu )/4}
-\Theta(-p_0)e^{-\beta(p_0-\mu )/4}}{\sqrt{e^{\beta(p_0-\mu )/2}+
e^{-\beta(p_0-\mu )/2}}}  \, \, .
\eqq
In this notation the parity odd part of $\Pi_{\mu \nu}$ becomes
\bqq
\Pi_{\mu \nu}^{odd}(p)&=& i e^2 \int \frac{d^3 k}{(2 \pi)^3}
 Tr[\gamma_{\mu} U_F(p+k)
\frac{\ps+\ks+m}{(p+k)^2-m^2}U_F(p+k)\dag \gamma_{\nu} U_F(k)\frac{(\ks
 +m)}{k^2-m^2}
U_F(k) ] \nonumber \\
&\simeq& -2me^2 \epsilon_{\mu \nu \rho} p^{\rho} \frac{\partial}{
\partial m^2}\int \frac{d^3 k}{(2 \pi)^3}
U_F(k) \frac{1}{k^2-m^2} U_F(k)\dag  \nonumber \\
&=& -2me^2 \epsilon_{\mu \nu \rho} p^{\rho} \frac{\partial}{\partial m^2}
\int \frac{d^3 k}{(2 \pi)^3} \left[ \frac{1}{k^2-m^2}+2\pi i \delta (k^2-m^2)
n_F(k) \right]  \, \, ,
\eqq
where the $\simeq$ means that we have expanded the integrand around
$p_{\mu}=0$. Collecting all terms we arrive at the following expression
\beq
\Pi_{\mu \nu}(p) = -\frac{ie^2}{4\pi}\frac{m}{|m|}\left[ \tanh (\beta
 (m+\mu))+\tanh
(\beta (m-\mu)) \right]+ O(p^2) \, \, . \label{CS}
\eeq
In the limit of zero temperature and zero chemical potential eq.(\ref{CS})
 reduces
to the result of \cite{pani}, \cite{aitch}, \cite{das}.

{\bf Plasma oscilations}

The application of a time dependent perturbation on a plasma sets up
oscillations in the system, which are of the form of
 longitudinal and transverse traveling waves. This is to be contrasted
 with the static case of screening discussed earlier, where the
introduction of a static charge in the plasma polarized the plasma
and screened the electric field.

The longitudinal wave is associated with compression and relaxation
of the plasma and its dispersion relation is naturally obtained from
the density-density correlation function, namely $\Pi_{00} $. The value
of the frequency as the momentum is taken to zero in the dispersion relation
yields what may be called the longitudinal plasmon mass. Thus the plasmon
is a long-wavelength collective effect in the plasma, embodying the
contributions due to the many-body effect. The oscillatory
electric and magnetic fields accelerate the electrons and positrons in the
plasma and this in turn means that it takes a finite amount of energy to
excite an oscillation with vanishing momentum. This is what renders the
plasmon massive.

The dispersion relation for the transverse oscillations is obtained,
correspondingly from the transverse part of the photon two point function.
The frequency at vanishing momentum is the transverse plasmon mass.
However, rotational invariance requires that at zero momentum the two plasmon
masses should be equal. This has been explicitly seen in 3+1 dimensions
\cite{weld}, \cite{kap}.
In what follows, we verify this equality for a planar plasma and find an
explicit expression for the plasmon mass.

The dispersion relations for the longitudinal and the transverse plasmons
are obtained respectively from
\beq
{\omega_{L,T}}^2={{\bf k}^2}+\Pi_{L,T}(\omega ,{\bf k})
\eeq
in the limit where $\vertical {\bf k }\vertical \ll  \omega $.
$\Pi_{L,T}$ have already been defined in earlier sections eq.(\ref{pi}). In
terms of
the components of $\Pi_{\mu \nu}$, they are given by
\beq
\Pi_L(\omega , {\bf k})=-{{k^2}\over{{\bf k}^2}}\Pi_{00}
\eeq
and
\beq
\Pi_T(\omega {\bf k})={1\over 2}(g^{ij}+{{k^i k^j}\over{{\bf k}^2}})
\Pi_{ij}(\omega ,{\bf k}) \, \, .
\eeq
Here, we choose to work with a neutral plasma (i.e., we set the chemical
potential $\mu =0$).
We expand $\Pi_L$ and $\Pi_T$ in powers of $ {\vertical {\bf p} \vertical}
/p_0 $ and in
the high temperature limit where we can drop the electron mass in comparison
with the temperature, we obtain
\beq
\Pi_L(\omega ,0)\simeq {{e^2}\over{2\pi }} \int_m^{\infty }dx\ n_F(x)
\eeq
and
\beq
\Pi_T(\omega ,0)\simeq {{e^2}\over{2\pi }}\int_m^{\infty }dx\ n_F(x),
\eeq
where $n_F(x)\equiv {1\over{e^{\beta \vertical x \vertical  }+1}}$.
The integral can be performed trivially and in the high temperature limit,
yields
\beq
m_T^2 = m_L^2 \simeq {{e^2}\over{2\pi }}T (\log 2)  \, \, .
\eeq
Comparing with what we had for the electric mass in the case of screening
in an earlier section eq.(\ref{elmass}), we see that
\beq
m_T^2 = m_L^2 ={1\over 2}m_{el}^2 \, \, .
\eeq
It is instructive to compare this with what is obtained in 3+1 dimensions.
There,
\beq
m_T^2 = m_L^2 = {1\over 3}m_{el}^2 \, \, .
\eeq

Thus we have, in this section explicitly shown the equality of the masses
of the two plasmons and have related this mass to the inverse of the
Debye screening length obtained earlier.

{\bf Radiative Fermion Mass }

It is well known \cite{red}  that a fermion mass term in 2+1 dimensional QED
(with
an odd number of flavors) breaks invariance under parity explicitly. This
results in the radiative generation of a C-S term for the gauge field. This
in turn provides a gauge-invariant mass for the gauge field. Here, we are
interested in the converse effect, namely, the radiative generation of
fermion mass. This already occurs at zero temperature and density. In
what follows, we investigate the effect of finite temperature and density
on this radiative mass.

For simplicity we assume a zero bare fermion mass. The radiative mass is
 computed from the self-energy correction graph using a standard procedure.
At zero temperature and density the leading order correction to the fermion
 mass is obtained from:
\beq
\Sigma (p)
=ie^2\int \val \gamma_{\nu} S_F (p-k) \gamma_{\mu} D^{\mu \nu}(k)   \, .
\eeq
 This corresponds to:
\beq
\Sigma (p)=
ie^2\int \val \frac{L_{\mu \nu }(p,k)}{(p-k)^2
[k^2-M^2]}[P^{\mu \nu }(k)+iM\epsilon^{\mu \nu \lambda }{{k_{\lambda }}
\over{k^2}}] \, ,
\label{self}
\eeq
where,
\beq
L_{\mu \nu }(p,k)=\gamma_{\mu }(p-k)_{\nu }+\gamma_{\nu }(p-k)_{\mu }
-g_{\mu \nu }(\ps -\ks )+i\epsilon_{\mu \nu \rho }(p-k)^{\rho }.
\label{sig}
\eeq
and $M$ is the topological photon mass.
The physical fermion mass is defined as the location of the pole in
the propagator. Expanding eq.(\ref{sig}) around $p=0$
 we find
\beq
m_{phys}={{e^2}\over{2\pi }}{M\over{\vertical M \vertical }}.
\eeq
At $T \ne 0$ and $\mu \ne 0$ however the situation is different. Since charge
conjugation
invariance is explicitly broken, one expects that in addition to the above
diagram
there is a contributing ``tadpole'' diagram. To ensure the overall
neutrality of the plasma, an uniform background is introduced, described
by a classical current $J_{\mu }$, which contributes a term
$J_{\mu }A^{\mu }$ to the Lagrangian. In 3+1 dimensions, the contribution
of the ``tadpole'' is infrared singular, as it involves the zero momentum
limit of the massless photon propagator. This singular contribution is
seen to be canceled by the background charge distribution chosen to
maintain the neutrality of the plasma. In 2+1 dimensions, the contribution
of the ``tadpole'' is not entirely infrared singular. The photon is now massive
and the  contribution has a finite part in addition to the singular parts.
However, the background distribution that cancels these infinities also
cancels this finite part, as we show below. Thus, even in 2+1 dimensions
where the photon acquires a topological mass, the ``tadpole'' does not
contribute to the fermionic two point function.
The second order contribution of the neutralizing external current to the
fermion propagator is given by
\beq
(-i)^2 \int d^dx d^dyJ^{\mu }(x)\langle T \bar{\psi }(y)\gamma^{\nu }\psi(y)
A_{\mu }(x)A_{\nu }(y)\rangle \, \, .
\eeq
If we choose the background as uniform, we may represent it as
$J^{\mu }(x)=ea^{\mu }$, where $a^{\mu }$ is a constant. Thus, the above
expression reduces to
\beq
-ea^{\mu }S_{F}(p)\gamma^{\nu }S_{F}(p)D_{\mu \nu }(0) \, \, ,
\eeq
which after truncating the external fermionic lines yields
\beq
-ea^{\mu }\gamma^{\nu }D_{\mu \nu }(0) \, \, .
\eeq
The ``tadpole'' diagram, on the other hand, gives a contribution of
\beq
-ie^2 \int {{d^dk}\over{(2\pi )^d}}tr[\gamma^{\mu }S_F(k)]\gamma^{\nu }
D_{\mu \nu }(0) \, \, .
\eeq
So, the background field required to nullify the ``tadpole'' contribution
can be read off to be
\beq
a^{\mu }=-ie\int {{d^dk}\over{(2\pi )^4}}tr[\gamma^{\mu }S_F(k)] \, \, .
\eeq
This can be easily seen to be proportional to the density of the plasma,
as it should be in order to render it neutral.
Thus,
\beq
a^{\mu }=eg^{\mu 0}\rho \, \, .
\eeq
The neutralizing background obviously vanishes when $\rho =0$, as the
plasma in that case contains equal numbers of positrons and electrons
and is neutral by itself.
This background current gets renormalized by radiative corrections to the
\lq \lq tadpole", which however do not contribute to the radiative fermionic
mass.

Let us now return to \ref{self}.
At finite temperature and
density, the self-energy it is given by
\beq
\Sigma(p)={{Me^2}\over{2{\pi }^2}}\int \val (p\cdot k-k^2)
[{{n_F(p_0-k_0)\delta((p-k)^2)}\over{k^2(k^2-M^2)}}-{{n_B(k_0)\delta(k^2
-M^2)}\over{k^2(k-p)^2}}] \, \, .
\eeq
We evaluate this on the mass-shell,$\ps =0$  and $p^2=0$.
A rather straightforward evaluation yields
\beq
m(T,\mu ,e^2,M)={{e^2}\over{2\pi }}{M\over{\vertical M \vertical }}
[1+{{2T(\log 2)}\over{\vertical
M \vertical}}+\frac{2T}{|M|} \log (\cosh (\frac{\mu}{2T}))
-{{2T}\over{\vertical M \vertical }}
ln(1-e^{-\vertical M \vertical /T})] \, \, .
\eeq
This result disagrees with that of ref.\cite{kao} in the overall sign of
the finite temperature corrections.
For the low temperature limit, we get
\beq
m \simeq {{e^2}\over{2\pi }}{M\over {\vertical M \vertical }}[1+{{T (\log
 2 )}
\over {\vertical M \vertical }}] \, \, .
\eeq
On the other hand, the result should not be extrapolated naively
to high temperatures. In this limit, the radiative mass grows
 uncontrolably
large and one is called upon to invoke self-consistently arguments.

{\bf Anomalous Magnetic Moment}

The anomalous magnetic moment of fermions is defined by using the Gordon
decomposition for the fermionic vector current. In 2+1 dimensions it has
the following form:
\begin{equation}
\bar{u}(p+q) \gamma_{\mu} u(p)=\frac{(2p+q)_{\mu}}{2m} \bar{u} (p+q)
u(p) +i \epsilon_{\mu \nu \lambda} \frac{q^{\lambda}}{2m} \bar{u}(p+q)
\gamma^{\nu} u(p) \, .
\end{equation}
It is easy to see from this expression that if we couple the fermions to an
external electromagnetic field the coupling of the second term describes
 the coupling to the magnetic field. Thus the magnetic moment is the
coefficient of $i \epsilon_{\mu \nu \lambda} \frac{p^{\nu} q^{\lambda}}{m}$.
 To lowest order in perturbation theory the only diagram that contributes
is the vertex correction diagram:
\bqq
\Lambda_{\mu}(p,q)&=&(-ie)^2\int \vol \bar{u}(p-q)\gamma_{\nu}S_{F}(p-k-q)
\gamma_{\mu}S_{F}(p-k)\gamma_{\lambda}u(p)D^{\nu \lambda}  \nonumber \\
&=&\Lambda^{0}_{\mu}(p,q)+ \Lambda^{\beta}_{\mu}(p,q)  \, \, ,
\eqq
where $\Lambda^{0}_{\mu}(p,q)$ , $\Lambda^{\beta}_{\mu}(p,q)$ are the zero
 and finite temperature vertex functions  respectively. The zero temperature
calculation has been done in \cite{kogan}, $\mu^{0}=\frac{1}{m}
\left[ \frac{1}{2}+\frac{
1}{\kappa} \right]$, where $\kappa=4 \pi M$.
The anomalous contribution to the magnetic moment  is extracted from the
real part of the three point function with on-shell electrons and vanishing
external spatial momenta\footnote{the alternate limit
i.e. taking $p_{\mu}=(0,\vec{p})$ and letting $\vec{p} \rightarrow 0$ leads
to different results , however this was shown recently [4] to be an artifact
of the perturbative expansion.}.
\bqq
Re\Lambda^{\beta}_{i}&=&\int \vol \,
 \frac{L_{i}(k,p,q) \delta(k^2-M^2)}{[(k-p)^2-m^2][(p-k-q)^2-m^2]}
n_b(k_0)+ \delta (k^2-m^2)n_F(k_0)  \nonumber \\
& &\left[ \frac{L_{i}(k+p,p,q)}{[
(k+q)^2-m^2][(k+p)^2-M^2]}+\frac{L_{i}(k+p-q,p,q)}{[(k-q)^2-
m^2][(k+p-q)^2-M^2]} \right]  \nonumber \\
& &- \left[\rm {same\, \,\, with  \,} M=0 \, \, \right]   \, \, ,
\eqq
where
\beq
L_{i}(k,p,q)=2[ -{\bf q}_m {\bf k}_j \ks \epsilon_i^{jm}+2i {\bf k \cdot
 p k}_i-i \qs
\ks {\bf k}_i-i({\bf 2p-q})_i k^2-2 p_0 {\bf q}_j {\bf k}_m \epsilon^{0jm}
 \gamma_{i}] \, \, .
\eeq
Using $q^2 \ll m, M$ we find
\bqq
Re \Lambda^{\beta}_i(p,q)&=&  \frac{p_0}{m} q_j \epsilon_{i}^{j0} \left[
\int_{M}^{\infty}
 \frac{dw}{2\pi} w^2 \frac{M^4+4 w^2 m^2}{[M^4-4w^2 m^2]^2} n_b(w)+
\right. \\ \nonumber
&+& \left.  \int_{m}^{\infty} \frac{dw}{4 \pi} \left( \frac{(w^2+m^2)(M^4+4m^2
w^2)
-8m^2 M^2 w^2}{[M^4-4m^2 w^2]^2}-\frac{w^2+m^2}{4m^2 w^2} \right) n_F(w)
\right] \, \, .
\eqq
A closed form expression for these integrals cannot be given in the general
case. However, in the limits of high and low temperature these integrals can
be reduced to simple analytic functions. In the low temperature limit
one can replace $n_B(w) \sim e^{-\beta w}$ and $n_F(w) \sim e^{-\beta w}$.
 We then find
\bqq
\Lambda^{\beta}(p,q)_i&=& q_j p_0 \epsilon_i^{\, j0} \mu (T) \nonumber \\
 & =&  p_0 q_j \epsilon_{i}^{j0} \frac{1}{32 m \pi} \left[
8m^2 \left( \frac{\beta M^3}{M^2-4m^2}-1 \right) e^{-\beta M}+
 \beta^2 M^4 \left( E_1(\mbox{ \small $\frac{\beta M(2m+M)}{2m}$})
e^{\frac{\beta M^2}{2m}}+
\right. \right. \nonumber  \\
& & \left.  E_1(\mbox{ \small $\frac{\beta M(2m-M)}{2m}$})
e^{-\frac{\beta M^2}{2m}}
\right)+
4m M^2 \beta \left( (E_1(\mbox{ \small $\frac{\beta M(2m+M)}{2m}$})
e^{\frac{\beta M^2}{2m}}-
 \right. \nonumber \\ & & \left. \left.
E_1(\mbox{ \small $\frac{\beta M(2m-M)}{2m}$})e^{-\frac{\beta M^2}{2m}}
 \right)
\right] \frac{1}{\beta m^4 }+\frac{1}{32 \pi} \left[-8\beta m^4 E_1(\beta m)
+\frac{16M^2 m^3}{2m^2+M^2} e^{-\beta m}
\right.  \nonumber \\
& & + \beta (M^2-2m^2)^2
(E_1(\mbox{ \small $\frac{\beta
(2m^2+M^2)}{2m}$})e^{\frac{\beta M^2}{2m}}+E_1(\mbox{ \small $\frac{\beta
(M^2-2m^2)}{2m}$})e^{\frac{- \beta M^2}{2m}}) \nonumber \\
& & \left. 4m (M^2-2m^2) (E_1(\mbox{ \small $\frac{\beta
(2m^2+M^2)}{2m}$})e^{\frac{\beta M^2}{2m}}-E_1(\mbox{ \small $\frac{\beta
(M^2-2m^2)}{2m}$})e^{\frac{- \beta M^2}{2m}}) \right] \frac{1}{m^4} \, ,
\eqq
where $E_1(x)=\int_{-\infty}^{x} \frac{e^t}{t} $ is the exponential
 integral.
Fig. \ref{fig5} shows the behavior of this function for small $T$.
This result together with
the calculation of the parity odd part of the photon propagator can be
used to
calculate the gyromagnetic ratio $g$ at finite temperature. At zero
temperature $g$ is defined by
 \[ \mu=\frac{g}{2m}S=\frac{g}{2m}(\frac{1}{2}+\frac{1}{\kappa}) \, \, . \]
 For anyons, i.e. arbitrary $\kappa$, $S$
, it has been observed \cite{nair} that $g$ is also $2$ and that this is
an exact result. To one loop this has been verified in perturbation theory.
 Here we confirm this result by taking  the limit $\mu, T \rightarrow 0$,
when $Re(\Lambda^{\beta}_i) \rightarrow 0$ exponentially.

{\bf Evaluating the chemical potential}

In order to write our results in terms of physical quantities one must
replace the lagrange multiplier  $\mu$  in all the physical quantities
that we have computed by the temperature dependent particle density
$\rho /e$. To do this we calculate $\rho$ to leading order in $e^2$.
\bqq
\frac{\rho}{e}=<\bar{\psi} \gamma_0 \psi > &=& \int \vol Tr(\gamma_0
(\ks +m))2 \pi \delta (k^2-m^2) n_F(k_0) \nonumber \\
&=& -\int_m^{\infty} \frac{dw}{2 \pi} \frac{1}{\beta}\left\{
\log \left[ \frac{1+e^{-\beta (w+\mu)}}{1+e^{-\beta (w-\mu)}}
\right] \right\}
\nonumber \\
 &=& \frac{1}{2 \pi} \left\{ \frac{\mbox{dilog} (1+e^{-\beta(m+\mu)})}{
m+\mu}-\frac{\mbox{dilog} (1+e^{-\beta(m-\mu)})}{
m-\mu} \right\} \, \, .
\eqq

{\bf Conclusions}

In this paper we have discussed various properties of a planar electron
positron plasma. Static properties like the screening length were
 calculated using
a real time formalism. Dynamic properties such as the plasmon frequency
were also calculated;  We find that  $(m_{pl}^{L})^2 =( m_{pl}^{T})^2=
\frac{1}{2}m_{el}^2$. The coefficient of the Chern-Simons term which is
unique to 2+1 dimensional gauge theories with explicitly broken parity
, was calculated for finite temperature and density. In the limit
of zero density the results agree with those of \cite{aitch}, \cite{pani}.
In this limit we have also computed the correction to the electron
magnetic moment
to one loop and extracted the value of the gyromagnetic ratio. The value
of $g$ goes smoothly to $2$ as $T \rightarrow 0$, thus agreeing with the
zero temperature result of \cite{kogan},\cite{nair}.

Even though we
have discussed relativistic systems one could also use the results for
a condensed  matter system where the quasi-particles have linear
dispersion relation. The extension of to the nonabelian case might be
interesting especially since some condensed matter systems can be
represented by quasi-particles with non-abelian interactions.

A further interesting feature of this system that one can address
fruitfully is the effect of dissipative processes. One can discuss
the viscosity of the plasma as well as the damping of the plasmon
modes due to dissipation. Work in this direction is in progress.

{\bf Acknowledgements}

We gratefully acknowledge fruitful discussions with  Bei-Lok Hu,
O.W. Greenberg and V.P. Nair. We thank V. P. Nair for a critical reading
of the manuscript and for making useful suggestions regarding its
improvement. We thank Reshmi Mukherjee for her invaluable assistance
in generating the figures. This work was partially supported by the
National Science Foundation.

\newpage
\begin{figure}
\centerline{\psfig{file=fig1pg.ps,height=6.5in,width=6.5in,bbllx=70pt,bblly=200pt,bburx=587pt,bbury=710pt,clip=.}}
\caption{The Contour of integration in the Complex Time Plane}
\label{fig1}
\end{figure}

\begin{figure}
\centerline{\psfig{file=fig2pg.ps,height=6.5in,width=6.5in,bbllx=70pt,bblly=200pt,bburx=587pt,bbury=710pt,clip=.}}
\caption{$m_{el}^2$ as a function of $T$ for $\mu=0.5$.}
\label{fig2}
\end{figure}

\begin{figure}
\centerline{\psfig{file=fig3pg.ps,height=6.5in,width=6.5in,bbllx=70pt,bblly=200pt,bburx=587pt,bbury=710pt,clip=.}}
\caption{$\frac{\tilde{\Pi}(0 ,{\bf 0})}{e^2}$ as a function of $T$ for
 $\mu=0.5$.}
\label{fig3}
\end{figure}

\begin{figure}
\centerline{\psfig{file=fig4pg.ps,height=6.5in,width=6.5in,bbllx=70pt,bblly=200pt,bburx=587pt,bbury=710pt,clip=.}}
\caption{Leading term in the expansion of $\frac{1}{e^2 \mu  (0,{\bf p})}-1
 $ in ${\bf p}$ as  a function of $T$ for $\mu=0.5$.}
\label{fig4}
\end{figure}


\begin{figure}
\centerline{\psfig{file=fig5pg.ps,height=6.5in,width=6.5in,bbllx=70pt,bblly=200pt,bburx=587pt,bbury=710pt,clip=.}}
\caption{The temperature dependent part of the leading term of the
anomalous magnetic moment of electrons ($\mu=0$). Here we take $m=1$
and $M=10$.}
\label{fig5}
\end{figure}

\end{document}